\documentclass[aps,prl,preprint,groupedaddress,showpacs]{revtex4}

\usepackage{graphics}
\usepackage{amssymb}

\begin{document}

\title{Some universal trends of the Mie(n,m) fluid thermodynamics }

\author{Pedro Orea, Yuri Reyes-Mercado, and Yurko Duda}

\affiliation{Programa de Ingenier\'{\i}a Molecular, Instituto Mexicano del Petr\'{o}leo,
%Eje Central L\'{a}zaro C\'{a}rdenas 152,
07730 M\'{e}xico D.F., M\'exico;  $Email: yduda@lycos.com$}

\begin{abstract}

\noindent By using canonical Monte Carlo simulation, the liquid-vapor phase diagram, surface tension, interface
width, and pressure for the Mie(n,m) model fluids are calculated for six pairs of parameters $m$ and $n$. It is
shown that after certain re-scaling of fluid density the corresponding states rule can be applied for the
calculations of the thermodynamic properties of the Mie model fluids, and for some real substances.

\end{abstract}

\maketitle

\maketitle

%\section{Introduction}

\noindent Among the intermolecular effective interaction potentials, the Lennard-Jones one is by far the most
widely used for approximating the physics of simple nonpolar molecules in all phases of matter
\cite{praus1,john01,john02,curil,okras,vega,alej1,galli1,dunikov01, szhou3, paricaud06,gloor05}. The
attractiveness of the LJ model is mainly due to its more convenient mathematical form than to its accuracy in
representing the properties of real fluids. Some modifications of the LJ potential, like exp-6
\cite{paricaud06,galli1} or the family of Mie(n,m) potentials \cite{lekker99,mansoori,nasrabad08}, have shown to
be useful for the description of thermodynamic and dynamic properties of some real substances. The Mie(n,m) pair
potential, which is just a general form of the LJ model, is defined as,

\begin{equation}
\label{Mie}
 u (r) =  \epsilon \Big( \frac{n}{n-m}\Big ) \Big(\frac{n}{m}\Big )^{m/(n-m)} \Big [\Big (\frac{1}{r}\Big )^n -
 \Big (\frac{1}{r} \Big )^m \Big ],
\end{equation}

\noindent where $r$ is the interparticle distance reduced by the particle diameter, which is chosen to be the
unit length, $\sigma =1$; $\epsilon$ is the well depth. The temperature of the system is defined as $T = k_BT/\epsilon$.\\

 Recently, both theory and molecular simulations have been used to compute the properties of the Mie(n,m) model fluids
\cite{dunikov01, hase98, lekker00, paricaud06, nasrabad08, pana96, jakse05, fpe8, okumura00, galli1, gloor05,
glasser}. The Mie fluid potential is often said to be short-ranged, however, for any finite system size, the
potential is not rigorously zero at a distance of the half box length where the potential is typically
truncated. Some approaches were proposed for dealing with the long-range tail of the potential
\cite{john01,gloor05}. Perhaps, due to the commonly used procedure of potential cut-off during molecular
simulations, there are lot of contradictions in the literature. As an example, the recent results of Okumura and
Yonezawa \cite{okumura00} have indicated that the coexistence curve for the Mie (n,6) models scaled by the
critical temperature and density practically coincides with each other, when $7<n<32$. Dunikov and co-workers
\cite{dunikov01} have also demonstrated that the coexistence and interfacial properties of the LJ fluid
approximately follows the corresponding state (CS) principle, if calculated with different cut-off's of the
interaction potential. Meanwhile, Galli\'{e}ro and co-workers \cite{galli1}, by studying modified forms of the
Mie(n,m) potential, found that no CS approach is possible between potentials having different repulsive
exponents as well as different functional forms.

It is well accepted that CS rule permits the prediction of unknown properties of many fluids from the known
properties of a few \cite{praus1,kisel06,guggen45,volker05,vekilov06}. Its application to the model potential
fluids allows to avoid the usually timeconsuming molecular simulations, and also makes possible the utilization
of some important developments in statistical mechanics which would otherwise be prohibited by computational
difficulties.  Firstly the CS rule was derived by van der Waals based on his well-known equation of state,

\begin{equation}
%\label{CSL} F \Big ( \frac{\rho}{\rho_c}, \frac{T}{T_c},
\label{CSL} F \Big( \rho_R, T_R, P_R \Big ) = 0,
\end{equation}

\noindent where the variables were reduced by their critical values, $\rho_R = \rho/\rho_c;$ (reduced number
density or inverse molar volume), $T_R=T/T_c$ (reduced temperature), and $P_R= P/P_c$ (reduced pressure). Later,
the CS principle was extended by introducing additional parameters to the study of various types of molecular
fluids \cite{praus1}. Thus, in general, macroscopic CS law states that all substances obey the same equation of
state in terms of the reduced variables, or, in other words, the state of a system may be described by any two
of the three variables: pressure, density, and temperature. \\

%It is well-known, that the pair potential between real particles is never truly conformal. This is a reason why
%different ways to generalize CS principle have been proposed \cite{noro00,foffi06,valde03}.

\noindent Very recently \cite{duda08}, we have studied the application of the CS principle for the hard sphere
attractive Yukawa (HAY) fluid. This model fluid is far from being conformal \cite{praus1,noro00}, neither its
reduced second virial coefficient $B_2^*$ evaluated at the critical temperature for various $\kappa$ (range of
attractive tail) is constant.  However, we have shown, that for different values of $\kappa$, HAY fluid obeys
the CS law for various thermodynamic properties. Unlike the studies of Refs.\cite{noro00,foffi06,lekker00} where
the approximation of the constant value of $B_2^*$ was used, we have proposed a new rescaling of fluid density,
that allowed us to obtain a relation like eq.(\ref{CSL}) for the HAY fluid. In Ref.\cite{duda08} we have
suggested that the same kind of rescaling may be universal for different pair potentials, what afterwards has
been partially confirmed \cite{fpe8}. The purpose of the present investigation is to show, by applying canonical
Monte Carlo (MC) simulation, that different thermodynamic properties of the Mie fluid obey the principle of the
corresponding states. Unlike the previous studies \cite{galli1,okumura00}, here we present the simulation data
not only for the vapor-liquid coexistence densities, but also results of the surface tension, interfacial width,
and supercritical pressure calculations.

\noindent In this work we study the systems with two particular cases of the potential (\ref{Mie}). Namely, the
one-parameter Mie(2m,m) model, in which increasing $m$ leads to a shorter ranged potential. Our calculations are
focused on three values of $m$: $7,~9$ and $12$. Another potential studied here is Mie(n,6). In this case, like
in Refs. \cite{pana96, okumura00, jakse05, galli1}, only the repulsive part is manipulated;  three potentials
with $n=12,~18$, and $32$ are considered. \\

\noindent The applied simulation method is the same that was used in our previous works\cite{rendon06, romero07,
duda08}. Briefly, the simulations of the vapor-liquid interfaces were performed in a parallelepiped cell with
sides $L_x=L_y=12$, and $L_z$ was chosen to be, at least, three times longer than $L_x$; the number of particles
inside the box was $N \geq 1500$, depending on the thermodynamic conditions considered. Periodic boundary
conditions were applied in all three directions. The pressure of the supercritical Mie(n,m) fluid was calculated
in a cubic simulation cell with $L_x=12$. We used $r_{cut}=5.5$ for Mie(12,6) and $r_{cut}=5.0$ for other
systems. As we verified, these values of $r_{cut}$ are sufficiently long to omit the influence of the
potential truncation on the averaged results presented below.\\

 Coexistence vapor $\rho_V$ and liquid  $\rho_L$ densities, and the interfacial width, $\delta$,
 were obtained at the end of each simulation run by fitting the density
 profiles, $\rho(z)$, to the following hyperbolic tangent function

\begin{equation}
\label{hyperb}
  \rho(z)=\frac12(\rho_L+\rho_V)-\frac12tanh \Big(\frac{
2(z-z_0)}{\delta} \Big),
\end{equation}

\noindent where $z_0$ is the position of the Gibbs dividing surface. The critical density and temperature for
these model fluids were calculated by using the rectilinear diameters law\cite{smit96} with the universal value
of $\beta=0.325$. The critical pressures were estimated on the base of the Clausius-Clapeyron equation
\cite{praus1}. Critical parameters are given in Table I.\\

%\section{Results and discussion}

\noindent  Fig. 1. depicts the reduced density, $\rho_R$, as a function of the reduced temperature, $T_R$. It is
clearly seen that the coexistence curves of Mie(2m,m) fluid with $m=7,~9$ and $12$, and Mie(n,6) fluid with
$n=12,~18$, and $32$ map onto a single master curve. As was mentioned above, the similar behavior for Mie(n,6)
and LJ fluids has been reported in Refs. \cite{okumura00,dunikov01}.  These results confirm that the
vapor-liquid phase diagrams of any Mie(n,m) potential obey the CS law, at
least in the range of considered parameters.\\

\noindent As usual \cite{romero07}, the interfacial tension is calculated by

\begin{equation}
\label{st} \gamma =\frac{L_z}2\Bigg\{\big<P_{zz}\big> - \frac 12\big[\big<P_{xx}\big> +
\big<P_{yy}\big>\big]\Bigg\},
\end{equation}

\noindent where, $P_{ii}$ are the components of the pressure tensor.

 Our results of the surface tension are presented in its reduced form \cite{guggen45},

\begin{equation}
\label{gamma} \gamma_r=\frac{\gamma}{\rho^{2/3}_cT_c}.
\end{equation}

\noindent The values of $\gamma_r$ as a function of the reduced temperature $T_R$ are plotted in Fig. 2. As
seen, the six sets of the surface tension data form a single master curve, which means that the surface tension
of Mie fluid also obeys the CS theorem. In other words, application of the corresponding state rule may avoid
the time consuming calculations of the surface tension for other pairs of parameters $m$ and $n$ if the critical
parameters are known. It is worth noting that the master curve in Fig.2 is slightly different from its
HAY fluid counterpart reported in Ref. \cite{duda08}, and is represented by the following empirical equation,\\

\begin{equation}
\label{gammaF} \gamma_r = 4.8 (1 - T_R)^{11/9},
\end{equation}

\noindent which resembles the expression proposed by Guggenheim to correlate the experimental data
\cite{guggen45}.

\noindent In many engineering problems (for example, for dispersant applications in petroleum and pharmaceutical
industries) it is of high importance to predict the vapor-liquid interfacial width as a function of temperature.
Since such calculations are not trivial due to the spacial fluctuations of the interfaces, application of the CS
principle might be of high utility. Since the value of surface tension is inversely proportional to the
interface width \cite{alej1,romero07}, it is natural to expect that the rescaling of $\delta$ must be like
Eq.(\ref{gamma}),

\begin{equation}
\label{delta}\delta_r=\frac{\delta}{\rho^{2/3}_cT_c}.
\end{equation}

\noindent In Fig. 3 the reduced widths of the vapor-liquid interface, $\delta_r$, as a function of $T_R$ are
shown. As seen, the interface width curves of the Mie fluids are almost overlapped with each other, taking into
account the error bars. It confirms that calculation of interfacial width can also be simplified by application
of the CS principle with rescaling (\ref{delta}). Moreover, our preliminary calculations of $\delta_r$ for some
parameters of HAY fluids indicate
that it is practically the same as the one presented in Fig. 3.\\

\noindent Finally, we analyze the reduced pressure $P_R$ as a function of reduced density $\rho_r$, defined as

\begin{equation}
\label{rho} \rho_r=\frac{\rho}{\rho^{2/3}_c}.
\end{equation}

\noindent Such rescaling of fluid density has been proposed and tentatively justified in our previous work
\cite{duda08}, where the extended corresponding states law has been applied for the
description of the  $PVT$ properties the HAY fluid.\\

\noindent The values of $P_R$ for three reduced temperatures are presented in Fig.4; as seen, the reduced
pressure data for all Mie fluid potentials in question match on the same master curves at each value of $T_R$.
Besides, the results of $P_R$ for the Mie(n,m) family fluids and the HAY fluid \cite{fpe8} practically coincide.
The best agreement is reached at the lowest temperature and densities; only slight discrepancy, almost within
the error bars, is observed at high fluid pressures, $P_R > 45$. The discrepancy is expected to be more
pronounced at higher reduced pressures. By using the critical parameters presented in Table II, we also show in
Fig. 4 the reduced pressure for the three real fluids, $Ar, N_2$, and $CO$. As expected \cite{nasrabad08}, the
pressure of argon is predicted quite well by the Mie potential model, while some deviation of $N_2$ and $CO$
pressures from the simulation data may be attributed to the slight non sphericity of these gas molecules.

%\section{Conclusions}

\noindent In summary, we have calculated the vapor-liquid equilibrium and interfacial properties of Mie(n,m)
fluids using canonical Monte Carlo simulations. Our results indicate that the coexistence densities and the
surface tension, as well as the width of the vapor-liquid interface obey the corresponding states principle. On
the base of new accurate critical parameters (evaluated from our MC simulations), it is found that the critical
compressibility factor is around $0.3$, which agrees well with experimental data for some real
substances\cite{guggen45,sigma}. The new density rescaling, $\rho_r=\rho/\rho^{2/3}_c$, is universal for the
calculation of the reduced pressure of Mie(n,m), HAY \cite{duda08}, and Sutherland fluids \cite{fpe8}. We expect
that the proposed new criteria could be of relevant help for testing new theoretical approaches for the
investigation of the model, as well as real fluid systems \cite{alej1,szhou3,paricaud06, nasrabad08, gloor05}.

\section{Acknowledgments}

\noindent The authors gratefully acknowledge the financial support of the Instituto Mexicano del Petr\'oleo,
under the projects D.31519/D.00480.\\

\newpage

\section*{  Figure Captions}

\noindent {\bf Fig. 1} Reduced vapor-liquid coexistence curves of the Mie(n,m) model fluid at different
interaction ranges. The solid line is just a guide for the eye, which is well described by the empirical
formulae proposed by Guggenheim \cite{guggen45} to build the coexistence curve for
argon. Error bars do not exceed the symbol size.\\

\noindent {\bf Fig. 2} Reduced surface tension $\gamma_r$ as a function of reduced temperature $T_R$ for the
same systems considered in Fig.1.\\

\noindent {\bf Fig. 3}  Reduced interfacial width, $\delta_r$, as
a function of $T_R$ for the same systems considered in Fig.1. The solid line is a fitting curve,
$\delta_r = \frac {\exp (T_R) }{(1-T_R)^{0.56}}$. \\

\noindent {\bf Fig. 4} Reduced pressure, $P_R$, as function of reduced density, $\rho_r$, at three reduced
temperatures $T_R=1.23, 1.52$, $1.92$, and $4.651$ (from bottom to top). Our data for Mie (12,6) at $T_R =
4.651$ coincide with the results of Johnson et al. \cite{john02}. Symbols and lines depict the simulation data
of Mie fluid, and experimental estimations for real gases\cite{db,sigma}, respectively. Solid line is a
guide for the eye. \\

\newpage

\centering \resizebox{0.463\textwidth}{0.34\textheight}{\includegraphics{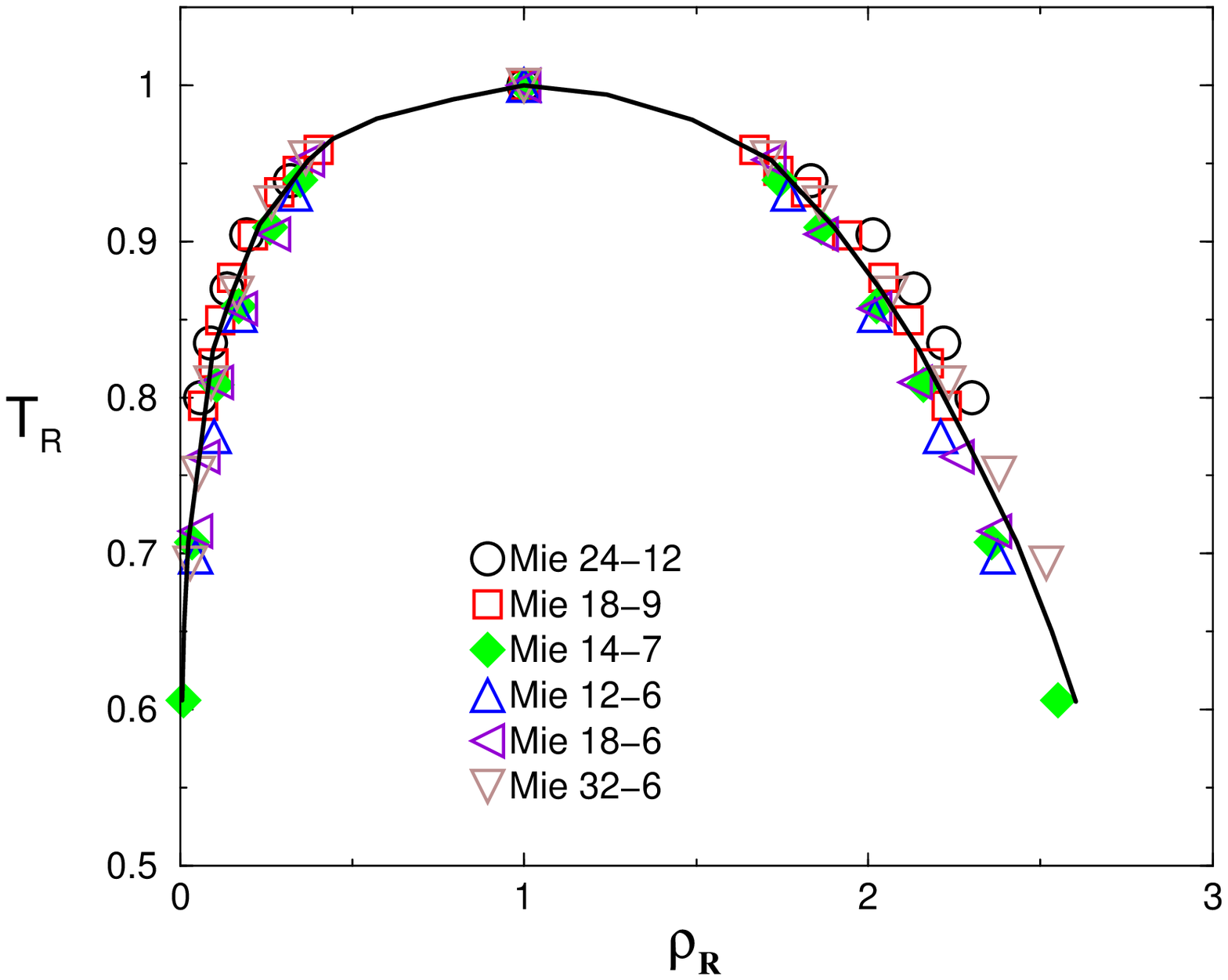}}

\vspace{1.0cm}

\centerline{Fig. 1}

\vspace{2.0cm}

\centering \resizebox{0.463\textwidth}{0.34\textheight}{\includegraphics{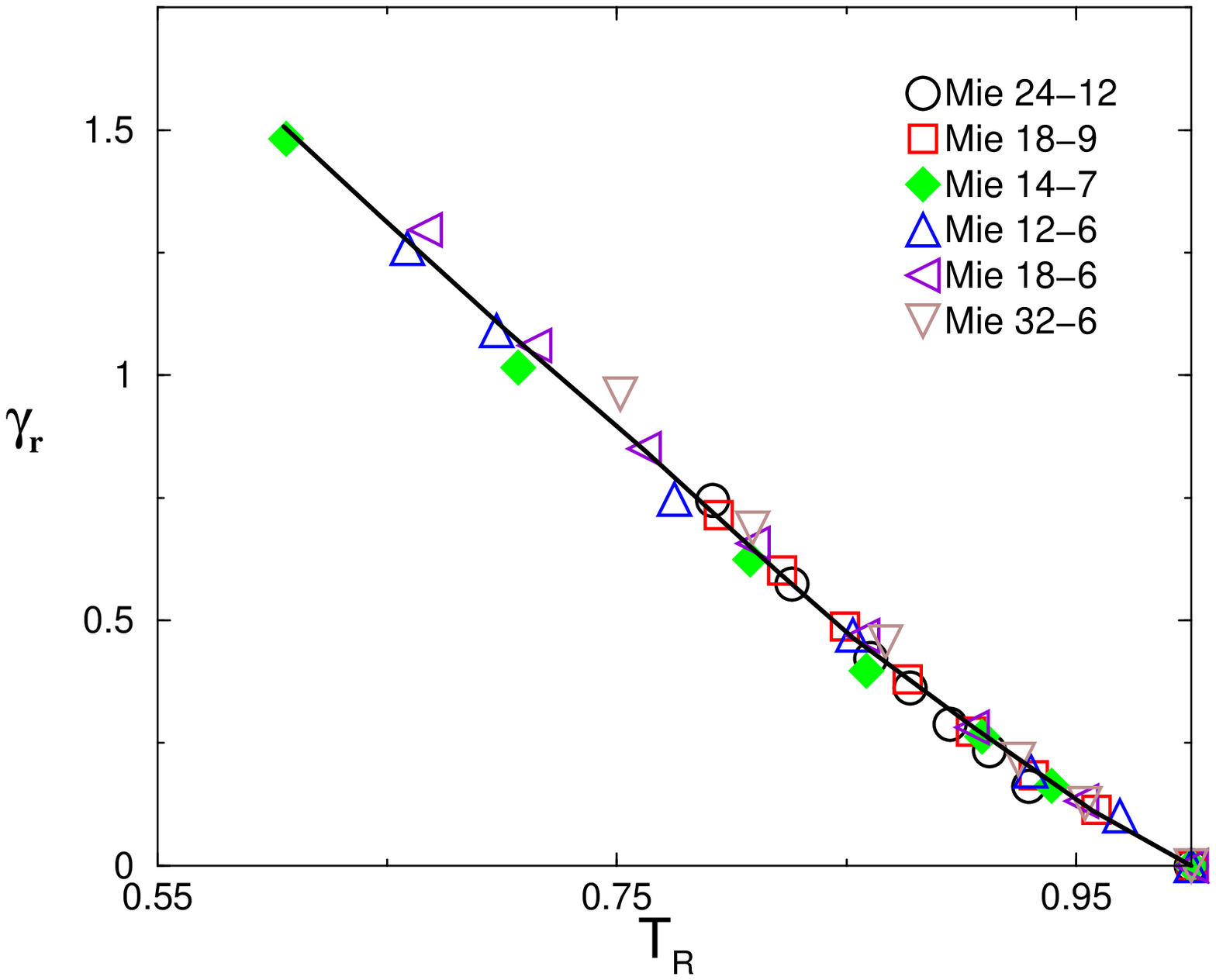}}

\vspace{1.0cm}

\centerline{Fig. 2}

\vspace{2.0cm}

\centering \resizebox{0.463\textwidth}{0.34\textheight}{\includegraphics{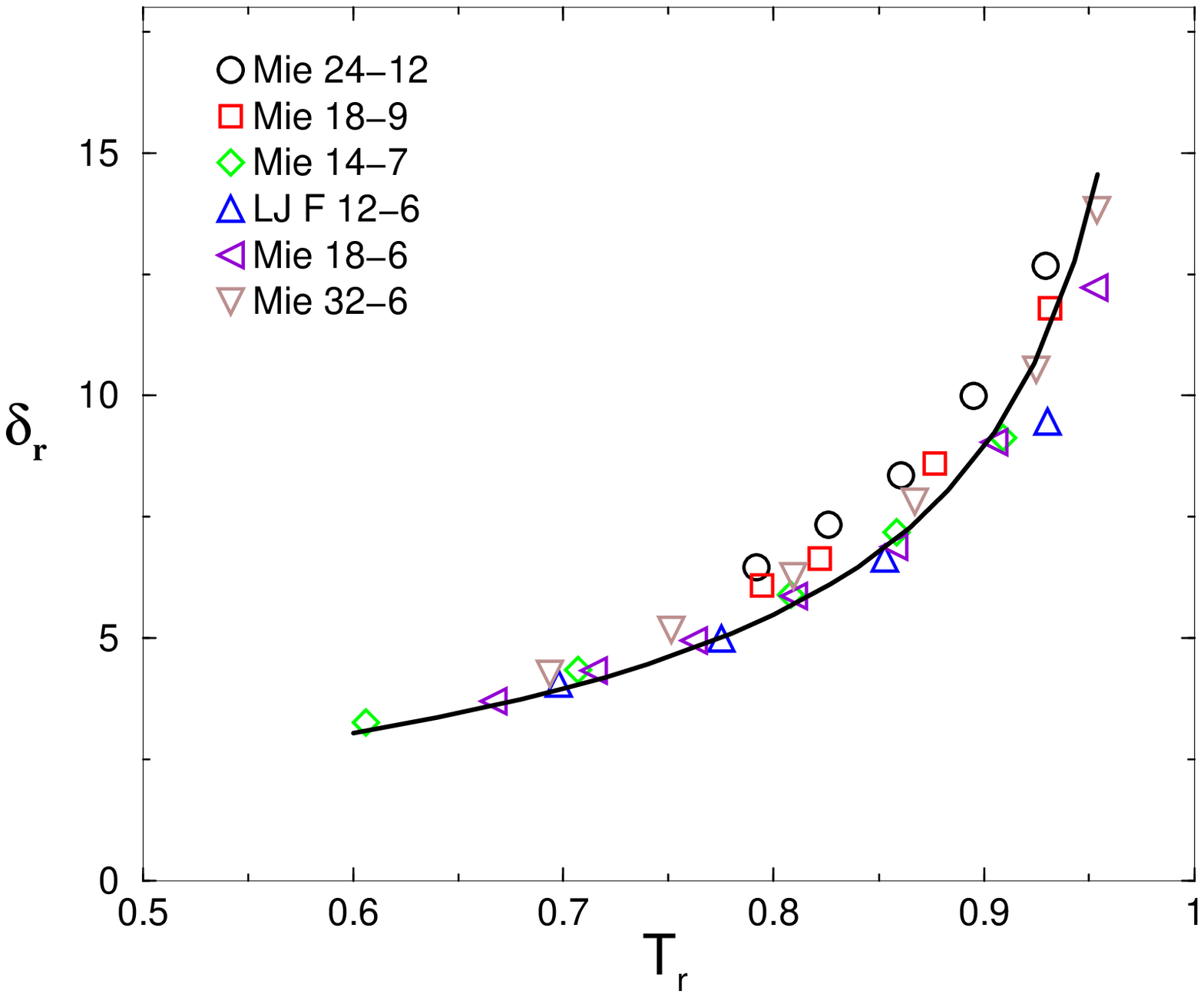}}

\vspace{1.0cm}

\centerline{Fig. 3}

\vspace{2.0cm}

\centering \resizebox{0.463\textwidth}{0.34\textheight}{\includegraphics{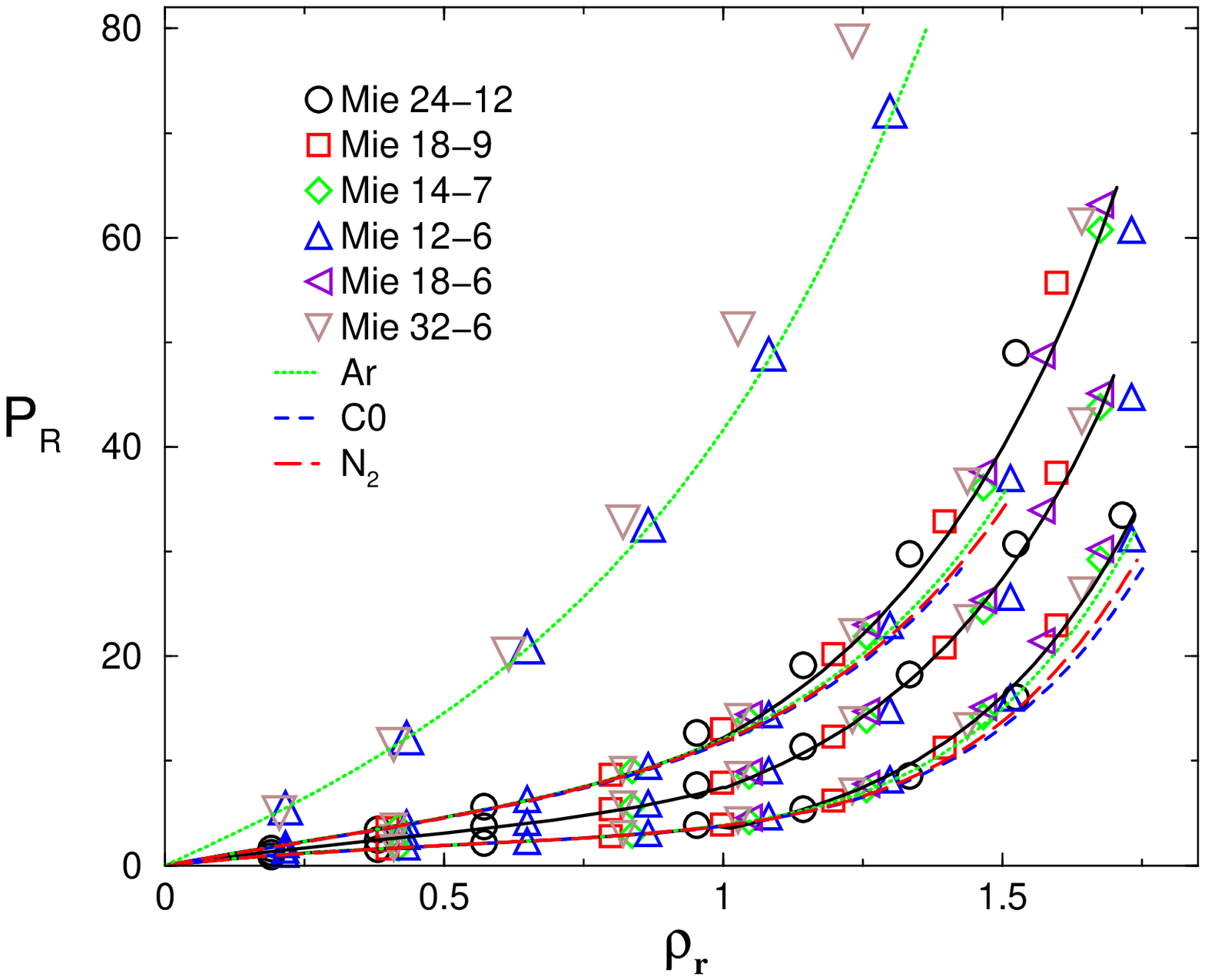}}

\vspace{1.0cm}

\centerline{Fig. 4}

\newpage

\begin{table}
\caption{Critical parameters of Mie fluids} \label{table1}

\begin{tabular}{ccccccc}
\hline\hline \hspace{1.5cm}$Systems$\hspace{1.5cm} &
$T_c$\hspace{1.5cm}  & $\rho_c$\hspace{1.5cm}
&  $P_c$\hspace{1.5cm}  &  $Z_c$ \hspace{1.0cm}  \\

\hline

 Mie(24-12)  & $0.575_{6}$  & $0.390_9$  &  $0.066_3$   &  $0.294_7$   \\
 Mie(18-9)   & $0.730_8$  & $0.360_9$  &  $0.078_2$   &  $0.297_8$   \\
 Mie(14-7)   & $0.990_7$  & $0.330_8$  &  $0.098_2$   &  $0.300_7$   \\
 Mie(12-6)   & $1.290_{9}$ & $0.314_6$  &  $0.118_{3}$  &  $0.291_8$   \\
 Mie(18-6)   & $1.050_6$  & $0.330_5$  &  $0.102_2$   &  $0.294_6$   \\
 Mie(32-6)   & $0.865_5$  & $0.340_7$  &  $0.090_3$   &  $0.306_5$   \\\hline \hline

\end{tabular}
\end{table}
.

 \vspace{2cm}

\begin{table}
\caption{Critical parameters \cite{db} and diameters \cite{sigma}  $\sigma$ of the three fluids analyzed in
Fig.4}.

\label{table2}

\begin{tabular}{ccccccc}
\hline\hline \hspace{0.2cm}$Substances$\hspace{0.2cm} & $T_c (K)$\hspace{0.7cm}  & $\rho_c$
(mol/L)\hspace{0.7cm}
&  $P_c$ (MPa)\hspace{0.7cm}  &  $\sigma$(pm)  \hspace{0.3cm}  \\

\hline

 $N_2$                   & $126.19$  & $11.1839$  &  $3.3958$   &  $375$   \\
 $Ar$                    & $150.69$  & $13.4074$  &  $4.863$   &  $342$   \\
 $CO$                    & $132.86$  & $10.85$  &  $3.4935$   &  $371$   \\\hline \hline

\end{tabular}
\end{table}

\end{document}